# Seasonality of hail outbreaks in the United States and its link to weather regimes

Matthew Graber,[a] Robert J. Trapp,[a] Zhuo Wang[a]

[a] *Department of Climate, Meteorology, & Atmospheric Sciences, University of Illinois Urbana-Champaign, Urbana, Illinois*

*Corresponding Author:* Matthew Graber, mgraber2@illinois.edu

ABSTRACT

Hailstorms in the United States produce immense economic losses and have an occurrence frequency that appears to have a long-term positive trend. Here we contribute an updated examination of such hail activity and the relevant environmental parameters over the period 1990-2024. We define a hail outbreak as any day with >6 significant hail reports (hail ≥ 2” diameter) and find a positive trend in hail outbreaks since 1990 (slope=0.8 days $yr^{-1}$) with the largest increases occurring in May and June and in the Southern Great Plains. Analysis of environmental parameters shows that the long-term increase in hail outbreaks correlates with a long-term increase in the significant hail parameter (SHIP, cc=0.67, p<0.01). The interannual variability and long-term trend of SHIP are driven both by thermodynamic and kinematic processes, though kinematic processes play a more important role than thermodynamic processes in the interannual variability of hail outbreaks. Using warm-season (April-July) weather regimes (WRs), we find that large-scale circulation modulates the interannual variability of hail outbreak frequency. An empirical model using WR frequency captures the interannual variability in warm-season hail outbreaks reasonably well (cc=0.38, p=0.03). Our study is the first to relate hail activity to WRs and presents a better understanding of the trend and year-to-year variability of hail outbreaks.

SIGNIFICANCE STATEMENT

Hailstorms produce major U.S. economic losses, yet the large-scale controls on high-impact events remain poorly understood. Herein, we define hail outbreaks as days with >6 significant hail reports (hail ≥ 2” diameter) and find that these events have increased since 1990, especially in April-July over the Southern Great Plains. Outbreak days account for 55% of hail-related losses and 68% of injuries, demonstrating their societal relevance. We link this increase to changes in both thermodynamic and kinematic environments, summarized by the significant hail parameter. We further show that weather regimes modulate the interannual variability of hail outbreaks, providing a potential framework for better hail prediction and understanding the mechanisms driving the trends in hail activity.

## 1. Introduction

Hailstorms produce major economic losses in the contiguous United States (CONUS) every year including damage to homes, vehicles, and crops (Battaglia et al. 2019; Allen and Tippett 2015).

Although hail results in relatively few fatalities, individual storms have produced damages exceeding a billion U.S. dollars (Brown et al. 2015), and hail is the leading contributor to insurance losses among severe weather hazards (Sander et al. 2013). Most hailstones are small, < 1” in diameter, however, hailstones as large as 6-8” have been observed; days with reports of ≥1” (severe) and ≥2” (significant) hail occur most frequently in the Southern Great Plains, particularly during the month of May (Allen et al. 2017). Additionally, the Great Plains experiences large interannual variability in hail occurrences, implying varying impact from year to year (Jeong et al. 2020, 2021). Despite these documented impacts, there is limited understanding regarding how hail events have varied over time and how this relates to the variability of the large-scale meteorological environment. Improving our understanding of these high-impact events is critical for assessing societal risk and informing stakeholders.

A common question among stakeholders and others regards the possibility of long-term trends in such hail events (Raupach et al. 2021). Allen and Tippett (2015) provide the most comprehensive analysis of CONUS hail trends to date. Although CONUS hail days exhibit minimal decadal variation (Allen and Tippett 2015), positive trends in the number of days with reports of hail of diameter ≥ 0.75”, ≥ 1”, and ≥ 2” exist from 1955 through the mid-1990s, consistent with increases in total hail reports (Allen and Tippett 2015). However, such positive trends vanish thereafter, at least until 2014. Updating the analysis of Allen and Tippett (2015) with an additional decade of data and extending it to specific seasons and regions would provide valuable insight and thus is pursued herein.

Not considered by Allen and Tippett (2015) was an examination of the frequency of days with multiple reports meeting these size thresholds, perhaps owing to the lack of a formal definition of a hail outbreak. Schlie et al. (2019) addressed this by introducing a radar-based outbreak definition using the maximum expected size of hail (MESH), a radar-derived estimate of the largest expected hailstone at the surface (Witt et al. 1998), and demonstrated a linear relationship between hail reports and MESH area. Using this metric, they defined a hail outbreak day as any day in which the MESH area of grid points expecting hail ≥ 0.75” exceeded the 90$^{th}$ percentile for 2000-2011 and found that such outbreak days increased over that period, particularly in June. Furthermore, they reported an increase in the number of hail swaths per outbreak day, indicating that hail on outbreak days has become more widespread.

The increase in hail swaths per outbreak day suggests that environments are becoming more capable of supporting frequent and intense thunderstorms within individual days. A fundamental, thermodynamic ingredient for thunderstorm formation and intensity is convective available potential energy (CAPE), which governs potential updraft strength through the relationship $w_{max} = \sqrt{2 \times \mathrm{CAPE}}$ (Holton and Hakim 2012) and therefore influences the potential for large hail production. However, while sufficient CAPE is necessary for large hail, excessively large CAPE may produce updrafts strong enough to eject hail out of the optimal growth zone before they can grow larger (Lin and Kumjian 2022), so the relationship between hail and CAPE is not a simple linear one. Consequently, relying solely on CAPE may overlook key processes that influence hail size. For example, recent work has found that wider updrafts (Homeyer et al. 2023) and corresponding large overshooting top depths (Christo et al. 2025) favor larger hail, indicating that additional factors modulate hail size (Nixon et al. 2023). Specifically, stronger vertical wind shear enhances updraft rotation, hail growth volume, and hailstone residence time, promoting larger hail in numerically simulated supercells (Dennis and Kumjian 2017). Likewise, observational studies link enhanced vertical wind shear to long-lived supercells capable of producing significant hail (Bunkers et al. 2006). Indeed, these findings highlight that both kinematic and thermodynamic factors influence hail production and likely their variability. Recent increasing tendencies in hail activity likely reflect a combination of these factors, and further knowledge is important for understanding the variability of hail activity.

A useful way to investigate these combined influences is through the significant hail parameter (SHIP), a composite parameter for hail-specific environments which was identified as one of the most important variables in the machine learning model for large-hail prediction developed by Czernecki et al. (2019). Tang et al. (2019) used SHIP for a CONUS study and found that there has been an increasing trend in large-hail environments east of the Rocky Mountains, suggesting that changes in the thermodynamic and kinematic environments favorable for hail may be contributing to the observed increases in hail days (Allen and Tippett 2015). Motivated by these findings, this study seeks to explore further relationships between hail outbreaks, SHIP, and their possible long-term trends.

This study additionally seeks to explore whether large-scale circulations can modulate hail outbreak frequency through their influence on SHIP and related environmental conditions. The plausibility of such an influence stems from the documented link between weather regimes (WRs) and the variability of tornado activity (Graber et al. 2025; Miller et al. 2020; Tippett et al. 2024). WRs are recurring and persistent large-scale circulation patterns representing a finite number of equilibrium states of the climate system (Charney and DeVore 1979; Hannachi et al. 2017; Michelangeli et al. 1995). In addition to tornadoes, WRs have also been used to investigate the variability of temperature and precipitation (Grams et al. 2017, 2020; Robertson and Ghil 1999; Schaller et al. 2018; Vigaud et al. 2018), but have not been used specifically for hail. Graber et al. (2025) developed an empirical model for tornado activity that employs WR frequency and persistence and found that WRs were able to help explain the interannual variability of warm-season tornado activity. A similar model framework using WR frequency is developed herein to assess whether variability in hail activity can be explained by the variability in WRs.

The remainder of this paper is organized as follows. Section 2 describes the data, environmental parameters analyzed, and the WR-based empirical model. Section 3 provides an updated analysis of hail trends, including hail outbreaks over seasons and CONUS regions. It also includes an analysis of important environmental drivers and WRs to investigate how the variability of hail activity depends on the environment and large-scale circulations. The paper culminates in Section 4 with a summary of the results and its implications for future work.

## 2. Methodology

### *a. Hail Observations*

Year-round hail reports from 1955-2024 were obtained from the NOAA Storm Prediction Center (SPC) severe weather database. The full period is used to provide climatological context; however, trend analyses and the development of the hail outbreak definition primarily focus on 1990-2024, corresponding to the modern reporting era including coverage by the WSR-88D network and an increase in the number of storm spotters, and enhancements in convective storm field research (Crum and Alberty 1993; Allen and Tippett 2015; Trapp 2013). Reports are georeferenced and also include information on time, date, and hailstone diameter. Known limitations in the use of hail reports include those due to population-dependent reporting, the 2010-change in the severe hail threshold, and issues in hail-size reporting more generally (Allen and Tippett 2015). Our focus on hail days attempts to alleviate these limitations. Specifically, severe hail days are defined as any day with at least one hail report either ≥ 0.75” (prior to 2010) or ≥ 1” (after 2010), consistent with the change in the definition of severe hail in the U.S. (Allen and Tippett 2015). Significant hails days are defined as any day with at least one hail report ≥ 2”. The definition of a hail outbreak is developed after observational trends are presented.

*b. Significant Hail Parameter*

Environmental variables required to compute SHIP include most-unstable CAPE (MUCAPE), specific humidity and temperature at pressure levels between 1000 and 700 hPa, 500-hPa temperature, and 10-m and 500-hPa winds, the data for which were obtained from the ERA5 reanalysis (Hersbach et al. 2020). SHIP is defined as:

$$\mathbf{SHIP} = \frac{[\mathbf{MUCAPE} \times \boldsymbol{q} \times \boldsymbol{\gamma} \times (-\boldsymbol{T}_{\mathbf{500}}) \times \boldsymbol{S}_{\mathbf{06}}]}{\mathbf{44,000,000}} \quad \mathbf{(1)}$$

To determine $q$, the mixing ratio of a most-unstable air parcel, as well as MUCAPE, equivalent potential temperature ($\theta_E$) was calculated at each pressure level within the lowest 300-hPa at each ERA5 grid point. The specific humidity at the level with the highest $\theta_E$ was selected to represent the mixing ratio of a most-unstable air parcel. The term $\gamma$ denotes the 700-500-hPa lapse rate calculated as the temperature difference between 700 and 500-hPa divided by the corresponding vertical depth, and $S_{06}$ represents the 0-6-km bulk wind shear estimated using 10-m and 500-hPa winds, respectively.

Mean SHIP was computed for each April-July by averaging daily maximum SHIP ≥ 1.0 over land grid points. A threshold of 1.0 was used to restrict the analysis to environments supportive of significant hail. The specific focus on April-July corresponds to season of peak hazardous convective weather in the CONUS.

*c. Weather Regimes*

For consistency with the hail-report and SHIP analyses, April-July WRs were analyzed over the period 1990-2024 using daily 500-hPa geopotential heights (500H) obtained from the ERA5 reanalysis (Hersbach et al. 2020). To identify these WRs, the seasonal cycle, defined as the long-term 500H mean on each calendar day, was removed from the daily 500H at each ERA5 grid point, as was the linear trend of seasonal (April-July) mean Northern Hemispheric 500H. An EOF filter was then applied for dimension reduction, retaining the first 8 EOF modes, which together explain ~90% of the total anomaly variance. The April-July WR patterns derived in Graber et al. (2025)

for 1960-2022 using K-means clustering were used as reference WR patterns. Each day was assigned to the WR whose reference pattern had the smallest Euclidean distance from the daily EOF-filtered 500H anomalies.

*d. Empirical Model*

An empirical model was employed to evaluate the relationship between the hail outbreak variability and WRs. Following the conceptual framework of the tornado-index of Graber et al. (2025), a hail index (HI) was defined as:

$$\mathbf{HI(t)} = \sum_{i=1}^{5} \mathbf{f(i,t)} \times \boldsymbol{P_i} \tag{2}$$

where $t$ indicates year, $i$ indicates WR, and $P_i$ denotes the hail outbreak probability for WR-i, defined as the number of hail outbreak days occurring in association with WR-i divided by the total number of WR-i days. The modeled hail index for year t, $HI(t)$, is computed by multiplying the frequency of each WR, $f(i,t)$, by $P_i$ and summing across all WRs. Spearman Rank correlations and associated p-values are used to assess agreement between the observed and modeled hail indices.

## 3. Results

*a. Hail Observations*

The time series of annual severe hail days exhibits an increasing trend through 2000, followed by a decreasing trend to the present day (Fig. 1a). When analyzed as a function of month, we see that hail day occurrence during boreal spring and summer months approaches saturation by the end of the record, effectively limiting further increases (Fig. S1). Interestingly, when analyzed over the period 2000-2024, cool-season months show large decreases in severe hail days, particularly September (Slope = -0.221 days yr$^{-1}$), October (-0.483 days yr$^{-1}$), November (-0.27 days yr$^{-1}$), and February (-0.202 days yr$^{-1}$) (Fig. S2). Moreover, significant decreases in severe hail *reports* since 2000 are exhibited in nearly all months (Fig. S3). The suggestion therefore is that the apparent decrease in hail activity over the past two decades has occurred across all seasons.

Trends in significant hail days are consistent with ≥ 2” hail day trends reported by Allen and Tippett (2015), and show an overall increase, with the largest positive trend occurring during 1980-2000 (Fig. 1a), but no statistically significant trend from 2000-2024 ($p=0.1$). Monthly analysis (Fig. S4) shows statistically significant positive trends of significant hail days since 1960 in every month except November, with the largest trend occurring in May (+0.15 days $yr^{-1}$). Localized increases have occurred across portions of the Great Plains, including Texas (+0.46 days $yr^{-1}$), Kansas (+0.25 days $yr^{-1}$), and Nebraska (+0.31 days $yr^{-1}$) (Fig. S5). In addition, May and June show significant positive trends of significant hail reports from 2000-2024 (Fig. S6).

Further analysis of significant hail reports shows that days both with >5 (+0.77 days $yr^{-1}$) and >10 (+0.42 days/year) reports exhibit significant positive trends from 1955-2024 (Fig. 1b). However, these time series show distinct increases beginning around 1990. In fact, during the interval 1990-2024, occurrences of days with >5 and >10 significant hail reports increase at a faster rate (+0.81 days $yr^{-1}$ and +0.64 days $yr^{-1}$, respectively) than during the interval 1955-1989 (+0.31 days $yr^{-1}$ and +0.10 days $yr^{-1}$, respectively). Although meteorological factors may play a role, the contrasting trends from 1990-2024 likely reflect improved hail detection associated with the WSR-88D deployment in the early-mid 1990s (Crum and Alberty 1993). The rise beginning around 1989 also suggests that increased NWS warning verification efforts may have contributed, paralleling improvements in EF/F-0 tornado reporting (Allen and Tippett 2015; Verbout et al. 2006). Accordingly, our subsequent trend analyses focus on the 1990-2024 period.

The thresholds of 5 and 10 significant hail reports were somewhat arbitrarily chosen for use in Fig. 1b, but a closer inspection of the report distribution from 1990-2024 shows that the 70$^{th}$ percentile of the number of daily significant hail reports corresponds to 6 reports (Fig. 1c). Herein we adopt the threshold of >6 significant hail reports to define a hail outbreak day. This also allow a focus on the events of highest impact: Using NCEI property and crop loss data, along with SPC injury data, hail outbreak days based on this threshold account for ~55% of hail-related losses (property + crop) and 68% of hail-related injuries since 1993 (Pisut 2026), despite representing only 15% of all severe hail days.

The annual number of hail outbreak days has increased significantly since 1990 (Fig. 1d, $p<0.001$, slope = +0.8 days $yr^{-1}$), despite the decreases in annual severe hail days and relatively steady significant hail days since 2000. The annual number of days with >4 (60$^{th}$ percentile) and >8 (80$^{th}$ percentile) significant hail reports have also increased significantly (Fig. S7), thus implying the robustness of this trend with the chosen threshold and percentile. This indicates that potential meteorological changes during this period have led to an increasing tendency in hail outbreaks.

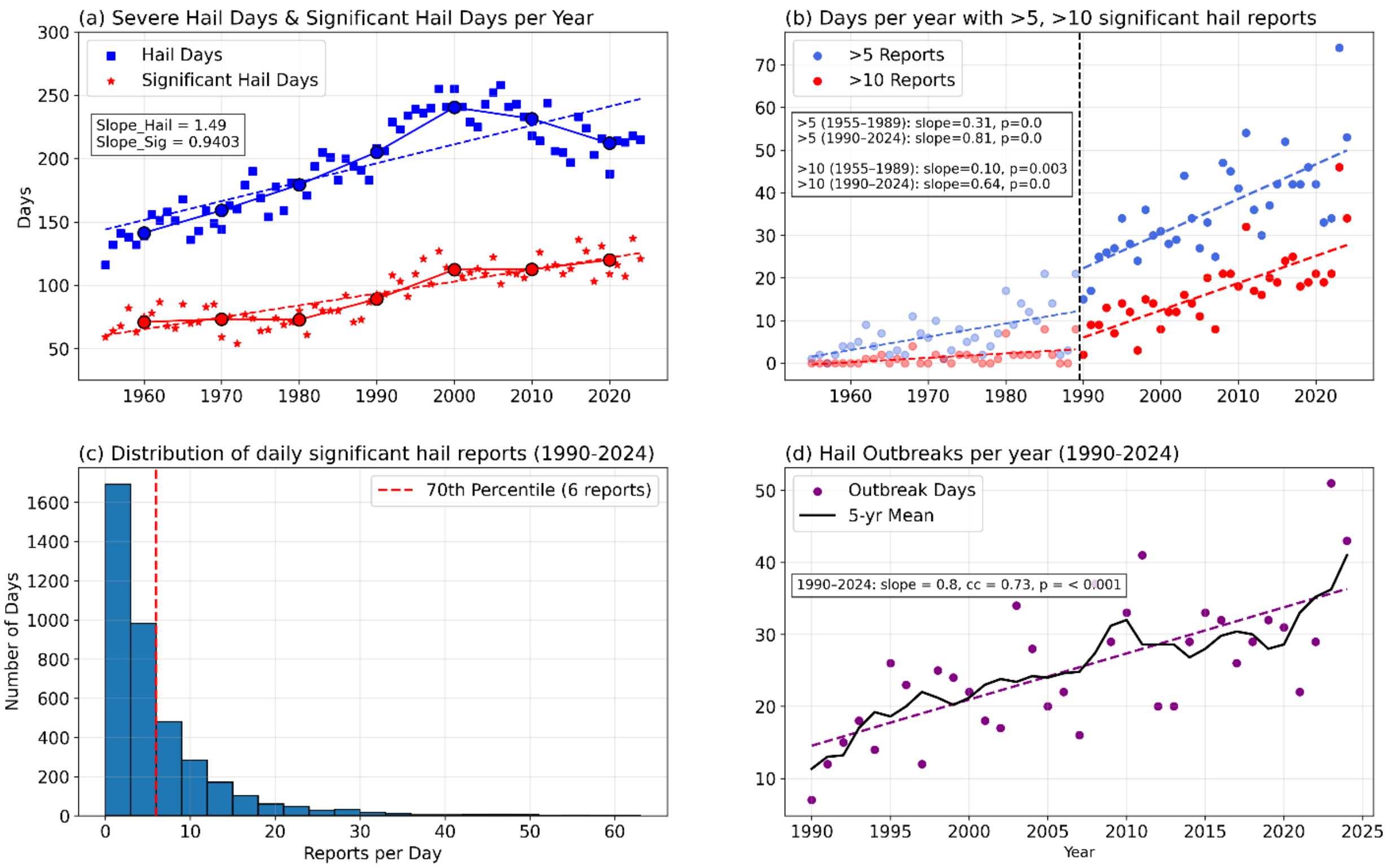

**Figure 1:** (a) Hail days (blue squares) and significant hail days (red stars) per year since 1955 with decadal averages (circles with black outline, solid lines) and linear regression slopes (dashed lines; day $yr^{-1}$). (b) Days per year with >5 (blue) and >10 (red) significant hail reports since 1955 with linear regression slopes (dashed lines; days $yr^{-1}$) and p-values. A black vertical line represents where the analyzed period (1990-2024) begins. (c) Distribution of significant hail report counts for every significant hail days since 1955 with labeled $80^{th}$ percentile for defining hail outbreaks. (d) Hail outbreaks per year since 1990 with 5-year running mean and linear regression statistics. All linear regression is done on raw data.

As a function of month, statistically significant positive trends in hail outbreaks are primarily confined to the boreal spring and summer months of April-July, with the largest trend occurring in May (Slope = +0.232 days $yr^{-1}$) (Fig. 2). The increasing trend in June hail outbreaks corresponds with an increase in June significant hail reports (Fig. S6) and is consistent with Schlie et al. (2019). August and September also have statistically significant increases (Slope = +0.078- and +0.035- days $yr^{-1}$, respectively) although the overall frequency of hail outbreaks during these months remains low. In contrast, the cool season contributes little to the annual trend in hail outbreaks, reflecting the rarity of significant hail and the infrequency of days exceeding 6 reports during these months (Fig. S3). Cool-season months have more days characterized by “high-shear, low-CAPE” conditions, particularly within the southeast U.S., which are generally less favorable for significant hail production, in contrast to similar "high-shear, high-CAPE" environments in the warm-season (Sherburn and Parker 2014).

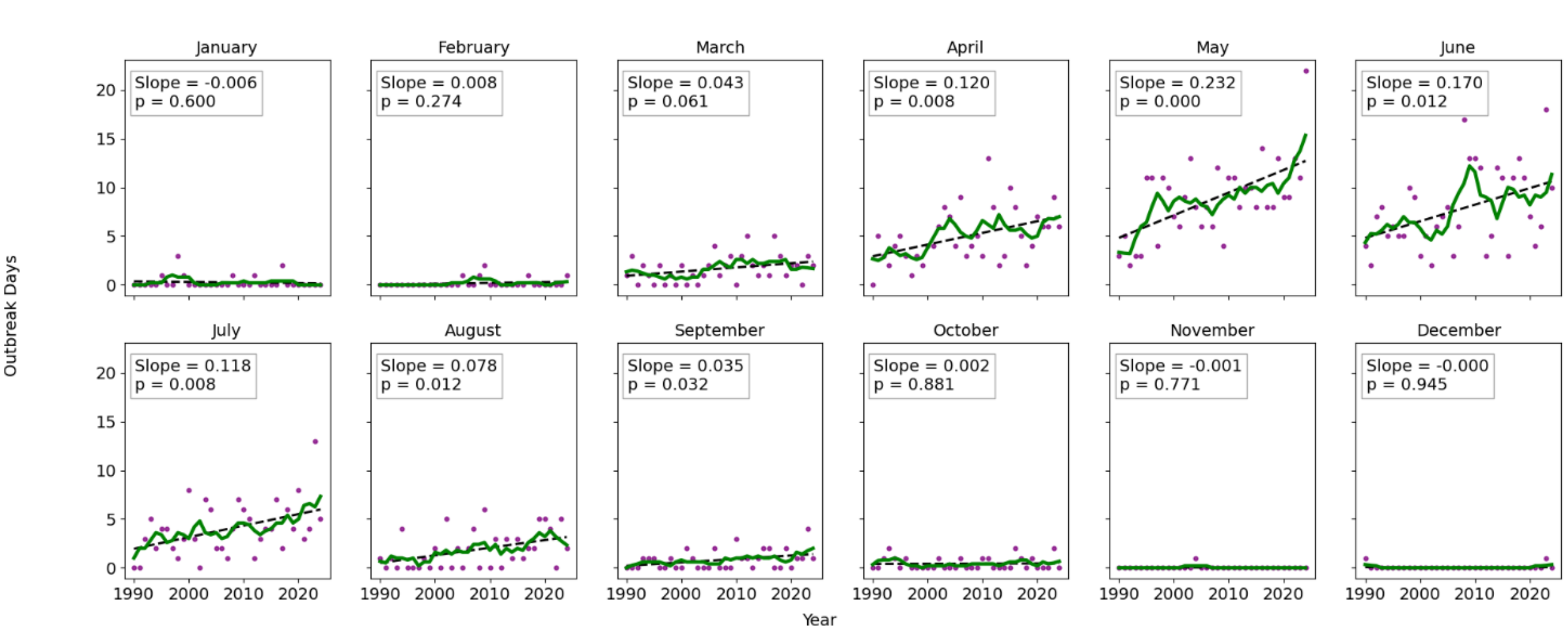

**Figure 2:** Hail outbreak days by month during 1990-2024 in CONUS (purple dots). Five-year running mean is indicated in green, and its linear trend is indicated by a dashed black line and summarized in each panel.

It is of additional interest to understand how these trends might be reflected in hail outbreak seasonality across the CONUS as well as within regions including the Northern Great Plains (NGP), Southern Great Plains (SGP), Midwest, and Southeast. This is analyzed by calculating hail outbreak probabilities on every calendar day using three periods spanning 1990-2024 (Fig. 3). Regions are defined following Moore (2018), except New Mexico is included in the SGP due to its similarity to Texas panhandle weather patterns (Shaw et al. 1997) (see Fig. S5).

Over these three periods, the CONUS experiences minimal shift from its late-May peak in hail outbreak probability, though the peak amplitude increases substantially, from 22% in 1990-2000 to 35% in 2012-2024 (Fig. 3a). Increases in the CONUS amplitude appear to be driven primarily by increases in the SGP (Fig. 3c). A broadening of the CONUS peak is due in part to outbreak increases in June and July within the NGP (peak probabilities from 4% in 1990-2000 to 10% in 2012-2024) and increases in April within the SGP (Fig. 2). The Midwest is the only region showing a shift in peak hail outbreak probability, moving from mid-May in the first period to mid-June in the third, though the probabilities are small; the probabilities are also rather small in the Southeast. Notably, this shift is absent in both the SGP or NGP. The stable hail seasonality differs from the seasonal shift observed with tornado activity (Graber et al. 2024; Gensini and Brooks 2018), suggesting the possibility of different climatological controls despite both hazards originating from severe thunderstorms.

All remaining analysis will focus on April-July (i.e. the warm-season), given its dominant contribution to long-term hail outbreaks trends.

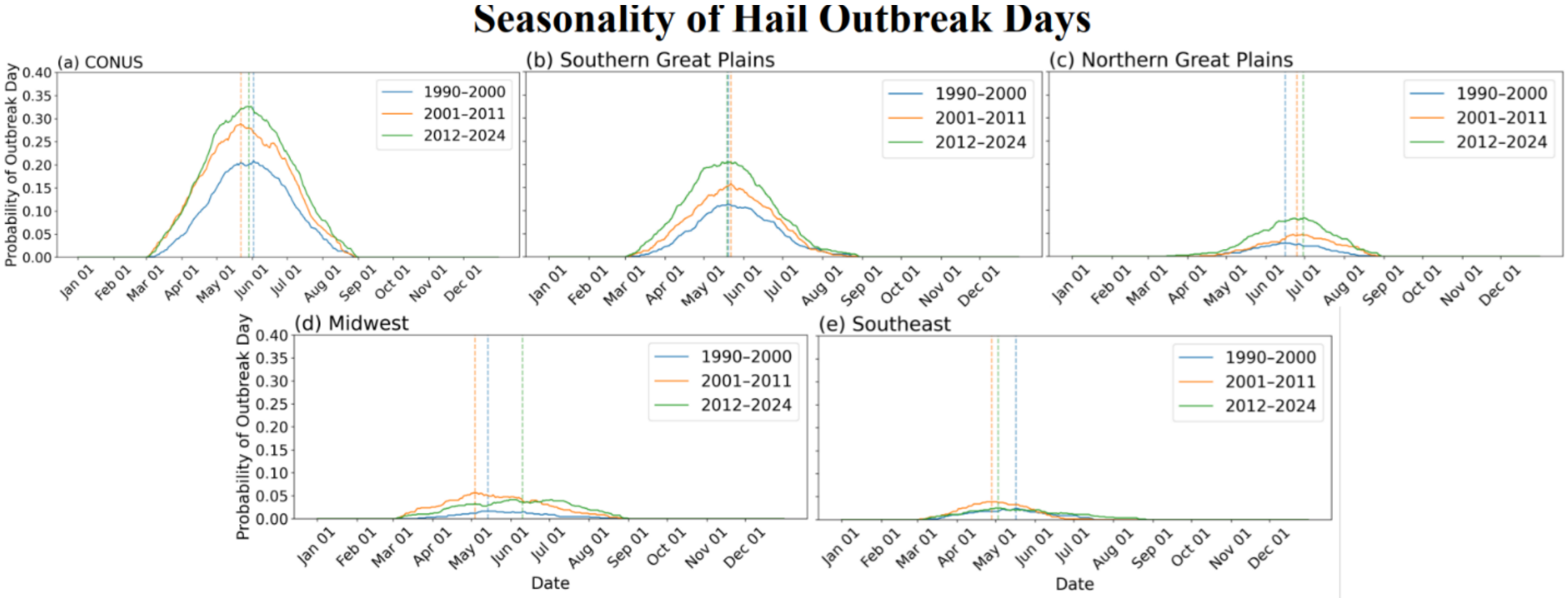

**Figure 3:** Seasonality of hail outbreak days in the (a) CONUS, (b) Northern Great Plains, (c) Southern Great Plains, (d) Midwest, and (e) Southeast depicted by calendar day probability that has been smoothed using a 61-day centered running mean for 1990-2000 (blue), 2001-2011 (orange), and 2012-2024 (green). Dash vertical lines represent the date of peak probability.

*b. Environmental Contributors*

Mean SHIP over each warm season during 1990-2024 is examined as a possible environmental contributor to warm-season hail outbreak variability (Fig. 4a). Warm-season mean SHIP exhibits a statistically significant positive trend over this period (slope=+0.0027 SHIP $yr^{-1}$, p=0.005), although the trend is weaker than that of the warm-season hail outbreaks (slope=+0.72 days $yr^{-1}$, p<0.001). The two time series are well correlated with each other (cc=0.68 p<0.001), suggesting that the increase in SHIP may contribute to the rise in hail outbreaks. In particular, the two time series display similar temporal variability from 2000-2024, though SHIP exhibits slightly larger fluctuations, possibly because SHIP tends to be overestimated in coastal regions of the CONUS (Prein and Holland 2018). An additional analysis using accumulated warm-season SHIP (following Gensini and Brooks 2018) yields comparable results (Fig. S8a) though the correlation with hail outbreaks is somewhat lower (cc=0.58, slope=+847 SHIP $yr^{-1}$).

Increases in mean SHIP suggest that rising environmental favorability has contributed to the upward trend in warm-season hail outbreaks. Since SHIP incorporates both thermodynamic and kinematic components (see Eq. 1), it is useful to determine which components matter most to SHIP's long-term trend and interannual variability. To examine the long-term trend of SHIP, raw time series of CAPE and $S_{06}$ were detrended at every grid point ($\mathrm{Trend} = \mathrm{slope} \times (\mathrm{year} - 1990)$), and the resulting detrended fields were separately substituted into Eq. 1 with the other parameters unchanged. CAPE and $S_{06}$ make comparable contributions to SHIP's long-term trend (Fig. 4b), with SHIP slope reduced to +0.001 $yr^{-1}$ with detrended CAPE and +0.0008 $yr^{-1}$ with detrended $S_{06}$, respectively. Although CAPE and $S_{06}$ appear to be the primary drivers of SHIP's long-term increase, the combined effects of the other SHIP variables are non-negligible. This is more evident with the accumulated SHIP analysis (Fig. S8b) where SHIP slopes remain strongly positive when CAPE or $S_{06}$ is detrended (+648 $yr^{-1}$ and +568 $yr^{-1}$, respectively).

To assess whether CAPE or $S_{06}$ contributes more to the interannual variability of SHIP, long-term seasonal means of CAPE and $S_{06}$ were calculated at every grid point and substituted into Eq. 1 to create SHIP with fixed CAPE and SHIP with fixed $S_{06}$ (Fig. 4c-d). The correlations between original SHIP and SHIP with fixed $S_{06}$, and between original SHIP and SHIP with fixed CAPE, are similar (cc=0.54 and cc=0.49, respectively). However, SHIP with fixed CAPE is much smaller in magnitude than the original SHIP. It implies that the nonlinear interactions between CAPE and the other variables involved in SHIP (i.e., $q, \gamma$ $and$ $T_{500}$) are non-negligible. This is physically reasonable given that these thermodynamic parameters are all tied to static stability and therefore may be closely related to the variations in CAPE. This interpretation is supported by examining the daily time series of SHIP, CAPE, and $S_{06}$ at individual grid points in central Texas and central Nebraska (for 1990; Fig. S9). CAPE makes a larger contribution than $S_{06}$ due to the large fluctuations of the daily SHIP time series. The SHIP time series with fixed $S_{06}$ closely follows the original SHIP (with cc=0.94 at both grid points), while fixing CAPE substantially reduces the magnitude of SHIP variability and also results in a smaller seasonal mean SHIP.

On the other hand, the correlation between warm-season hail outbreaks and SHIP with fixed $S_{06}$ is much lower (cc=0.15, Fig. 4d) than the correlation between hail outbreaks and SHIP with fixed CAPE (cc=0.53). These results suggest that kinematic variability associated with $S_{06}$ is the primary driver of the year-to-year hail outbreak variability, although CAPE still plays a non-negligible role. A similar interpretation holds when using seasonal accumulated SHIP (Fig. S8d). Overall, these findings indicate that both thermodynamic and kinematic processes contribute to the long-term trend and variability of SHIP, but kinematic processes appear to play a more important role than thermodynamic processes in the interannual variability of hail outbreaks.

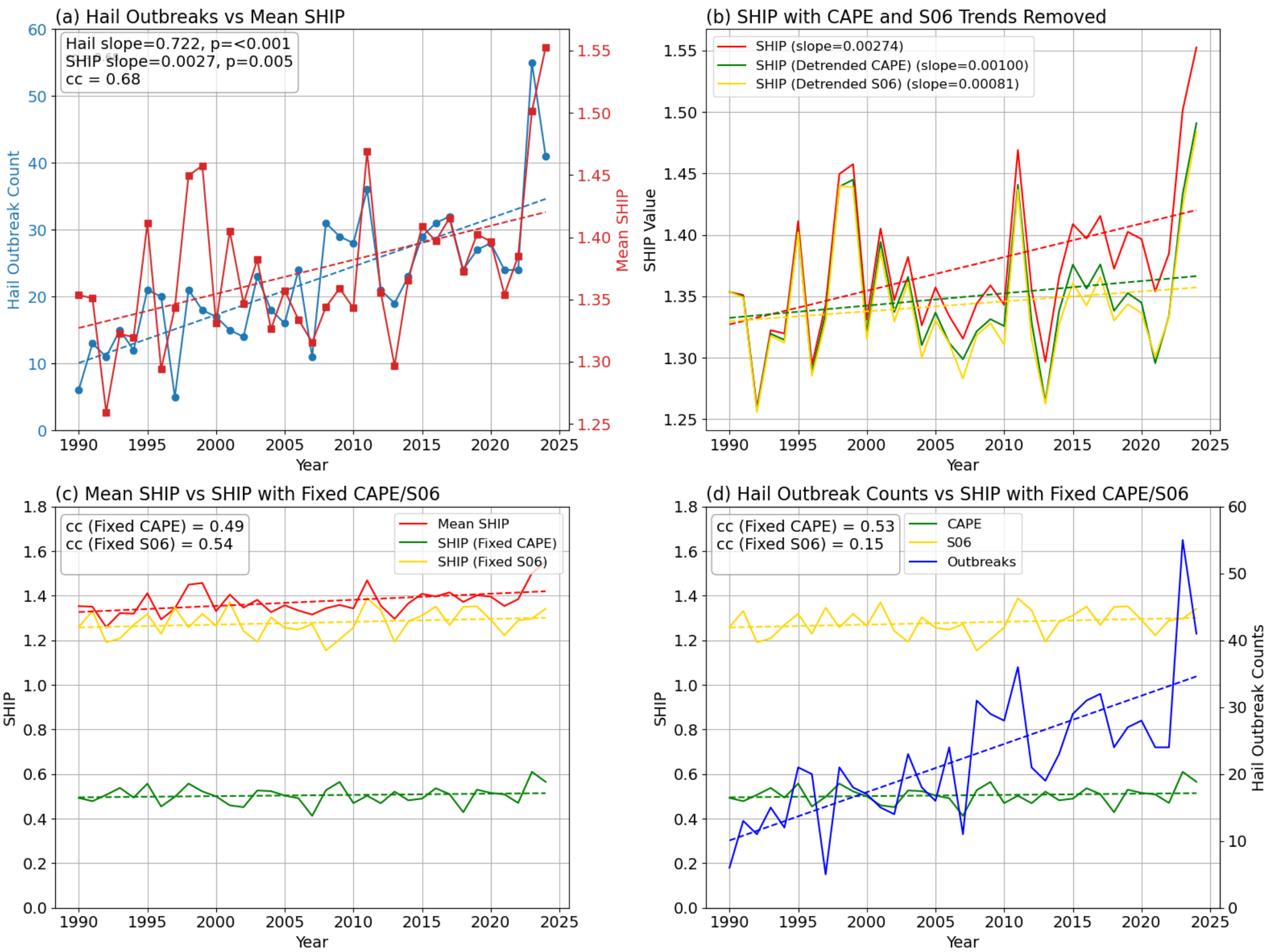

**Figure 4:** (a) Warm Season hail outbreak days (red) and warm season mean SHIP (blue; where SHIP ≥ 1.0 over land grid points only) from 1990-2024 with linear regression statistics for each time series and the correlation coefficient between both time series. (b) Time series of SHIP (red), SHIP with CAPE long-term trend removed, and SHIP with $S_{06}$ long-term trend removed. (c) Warm-season mean observed SHIP (red), and SHIP with fixed CAPE (green) and $S_{06}$ (yellow). Correlation coefficients are between the observed SHIP time series and the CAPE and $S_{06}$ time series. (d) Warm-season hail outbreaks (blue) vs SHIP with fixed CAPE (green) and $S_{06}$ (yellow) and corresponding correlation coefficients.

A clear geospatial relationship between warm-season SHIP and hail outbreak probability during 1990-2024 is revealed in Fig. 5f. Motivated by the results of Graber et al. (2025), we next investigate whether we can link the regional patterns of SHIP, and consequently of hail outbreaks, to WRs. April-July WR 500H patterns ~~are~~ shown in Fig. 5 ordered in decreasing frequency. Potential links to hail outbreaks are assessed using composite SHIP anomalies.

Relationships between WRs and hail activity are broadly similar to those for tornado activity (Graber et al. 2025), although hail outbreak probability anomalies are generally smaller in magnitude, indicating weaker, but still significant, modulation. Specifically, WR-A favors anomalously low SHIP over the entirety of the central CONUS corresponding to reduced hail outbreak probability anomalies. In contrast, WR-B features anomalously positive SHIP across the central CONUS, associated with enhanced hail outbreak probability anomalies. WR-C favors anomalously positive SHIP over the NGP, corresponding with less coherent, enhanced hail outbreak probability anomalies. Interestingly, hail outbreak probability anomalies over eastern Montana are stronger than their tornado counterparts in Graber et al. (2025), which is one of the few instances where hail probability anomalies exceed tornado probability anomalies. This may suggest WR-C is relatively more favorable for hail than tornadoes. WR-D favors anomalously positive SHIP over the SGP and NGP, corresponding to increased hail outbreak probability anomalies. A small portion of anomalously low SHIP over the east coast is associated with anomalously low hail outbreak probability anomalies. WR-E favors anomalously low SHIP anomalies over the NGP, which is associated with anomalously low hail outbreak probability anomalies. Anomalously positive hail outbreak probability anomalies over the Southeast are consistent with tornado activity (Graber et al. 2025). Of note is that anomalously positive hail outbreak anomalies exist over portions of the SGP as well despite the lack of anomalous SHIP, which may be partially explained by regions with lower climatological hail outbreak probabilities. Additionally, WR-E has the lowest long-term frequency (10.3%), resulting in the smallest sample size.

Positive and negative SHIP anomalies across all five WRs correspond well with positive and negative anomalies of CAPE and VWS in Graber et al. (2025). For example, the anomalous 500H high over the east coast and the anomalous 500H low over the central CONUS in WR-B increase vertical shear and enhance CAPE by facilitating moisture and warm-air advection from the Gulf of Mexico (Graber et al. 2025).

Thus, the relationship between the five WRs and hail-supportive environmental conditions, and the corresponding spatial alignment with the hail outbreak probability anomalies, appears to indicate that the variability of WRs plays a role in modulating hail outbreaks.

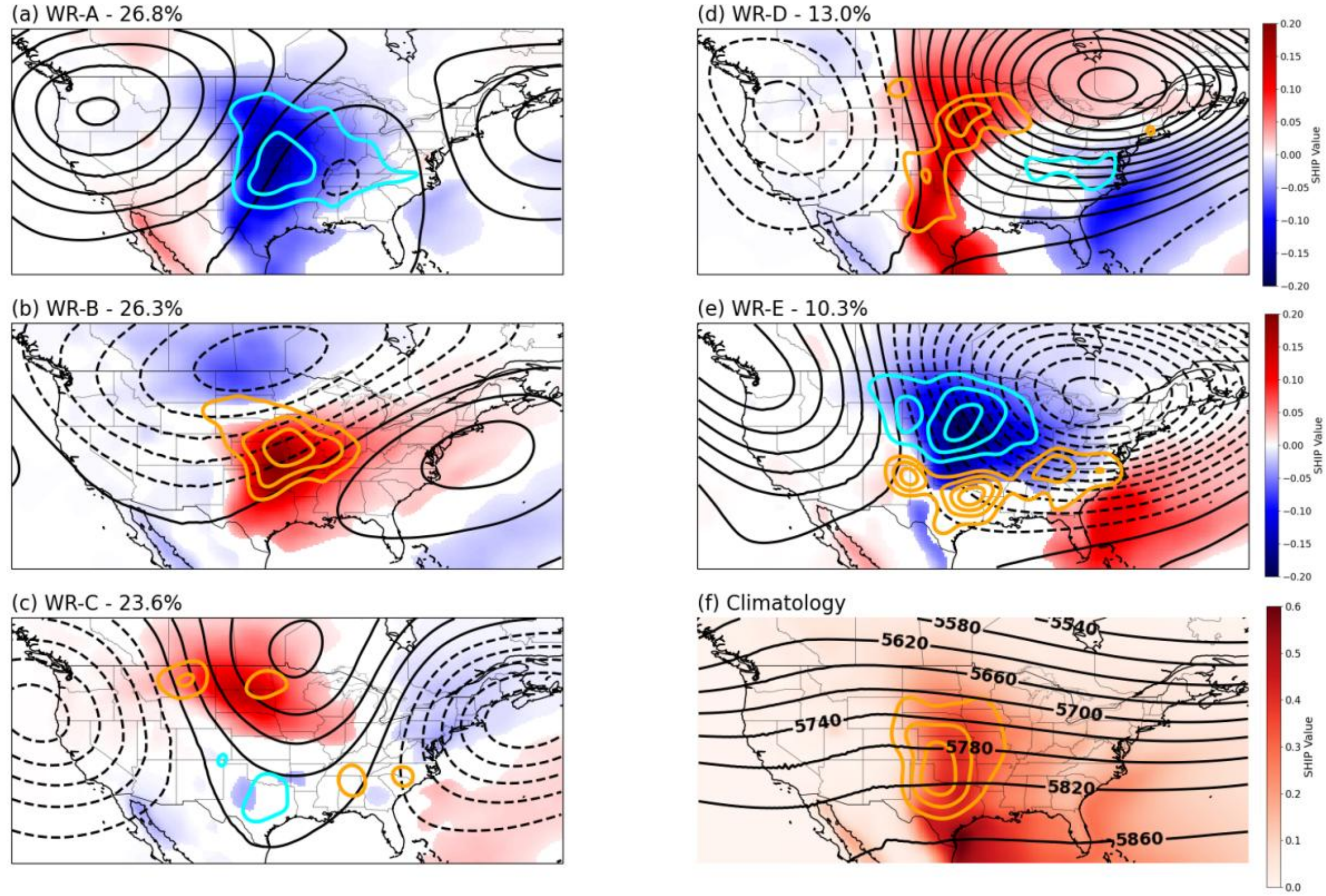

**Figure 5:** (a-e) The uniform-filtered, statistically significant composite anomalies of daily maximum SHIP (shading) at each 2° x 2° grid point (red/blue filled contours) for each WR with Gaussian-smoothed hail outbreak probability anomalies (contour intervals: $\pm 20$ %: orange and cyan colors represent positive and negative anomalies, respectively) and 500H for each WR (black contours: $\pm 10$ m). Significance ($p \leq 0.05$) is tested at each grid point using a one-sample, two-sided t test to mask out non-significant SHIP anomalies. (f) Climatology of daily maximum SHIP (shading), 500H (black contours), and hail outbreak probability (orange contours). Climatological probabilities of at least one report contributing to a hail outbreak (days with > 6 significant hail reports) are shown at each grid point with contour intervals of +0.0002.

To examine whether WRs can help explain the trend and interannual variability of hail outbreaks, we implement an empirical model based on WR frequency following Eq. (2). Fig. 6a shows CONUS hail outbreak probability anomalies for each WR, and Fig. 6b shows the corresponding empirical model. WR-A, B, and D exert the strongest modulation of hail outbreaks given their significant probability anomalies. WR-A is unfavorable for hail outbreaks consistent with the anomalously negative hail outbreak anomalies over the central CONUS (Fig. 5a). In contrast, WR-B and WR-D are favorable for hail outbreaks consistent with the anomalously positive hail outbreak anomalies over the central CONUS (Fig. 5b&d). WR-C and WR-E have insignificant CONUS hail outbreak probability anomalies consistent with scattered weak anomalies in WR-C (Fig. 5c) and the near-cancellation of anomalies in WR-E with positive anomalies in the SGP and negative anomalies in the NGP (Fig. 5e).

The empirical model fails to capture the observed positive trend in hail outbreaks. This suggests that the trend in hail outbreaks or SHIP (Fig. 4) is not due to trends in WR frequency but rather may be due to thermodynamic changes, such as increased CAPE (see Fig. 6 in Graber et al. 2025). Further investigation of such thermodynamic changes and underlying drivers (i.e., internal climate variability vs. external forcing) will help better understand the variability of hail activity but is beyond the scope of the current study. Nevertheless, the empirical model reasonably captures interannual variability in hail outbreaks and is significantly correlated with the detrended observed time series of hail outbreaks (Fig. 6b; cc=0.38, p=0.03). In particular, the model captures the frequent hail outbreaks in 2008 and 2011, which cannot be explained by SHIP (Fig. 4). This reflects the fact that WRs modulate CAPE and $S_{06}$ over the CONUS (Graber et al. 2025), which gives the variability of WR frequency a strong role in shaping interannual fluctuations of hail. Overall, Figure 6b suggests that dynamic changes in atmospheric circulations related to WRs play an important role in the interannual variability of hail activity, consistent with the discussion of Fig. 4.

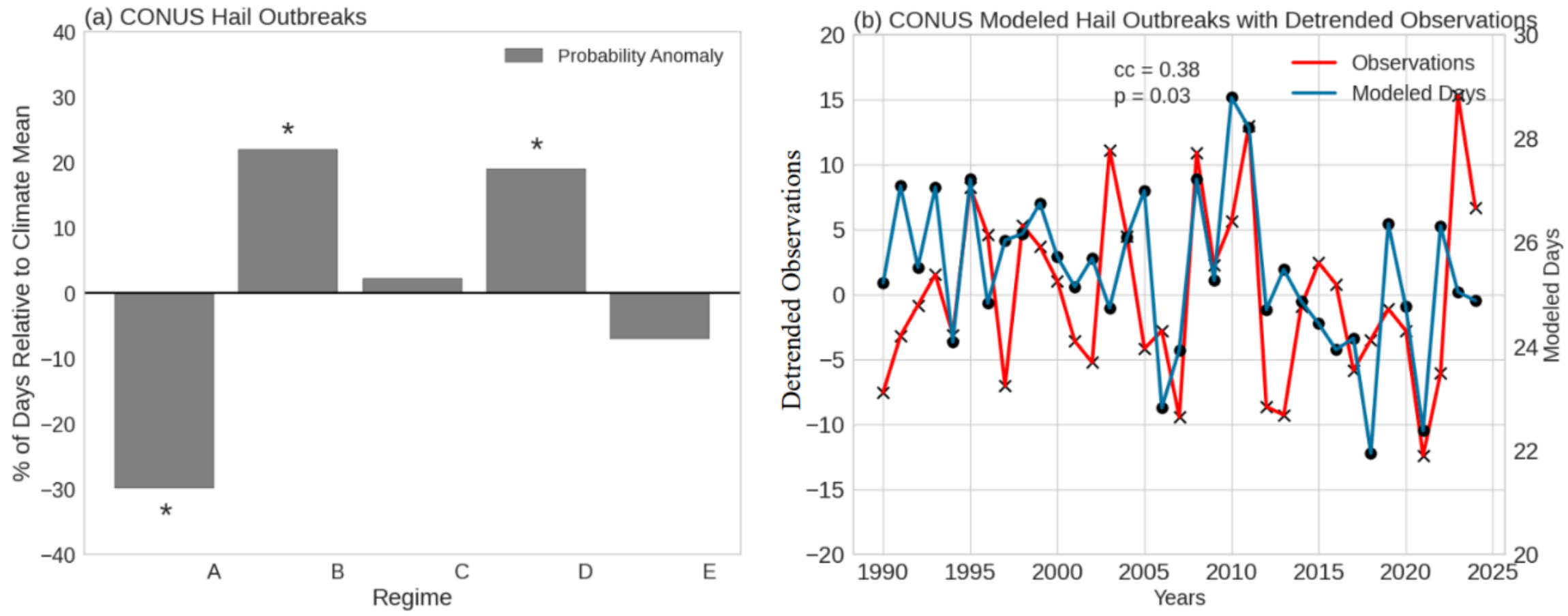

**Figure 6:** (a) Hail outbreak (days with >6 significant hail reports) probability anomalies for each WR. Anomalies above the 95% confidence interval based on a Monte Carlo test with 10,000 resamples are regarded as significant and are marked with an asterisk. (b) Empirically modeled hail outbreaks per year (blue curve with circles) overlaid with detrended observed hail outbreak days (red curve with x's) with Spearman rank correlation coefficient and p-value.

## 4. Summary

An examination of U.S. hail activity and the relevant environment conditions over the period 1990-2024 is presented. Despite overall decreases in severe hail days and no clear trend in significant hail days since 2000, hail outbreaks – herein, any day with >6 significant hail reports – have exhibited a significant increasing tendency since 1990. Spatial and temporal disaggregation shows that this annual tendency is driven by significant increases in hail outbreaks during boreal spring and summer, primarily over the SGP. Interestingly, the seasonality for hail outbreaks is relatively stable over 1990-2024. This differs from seasonality shifts observed for tornadoes (Graber et al. 2024; Brooks et al. 2014), suggesting the possibility that different climatological controls may have larger roles in modulating each hazard, but this goes beyond the scope of this study.

Mean SHIP over each warm season during 1990-2024 was examined as a bulk environmental parameter for warm-season hail outbreak variability. It exhibits a significant increasing trend from 1990-2024 and is highly correlated with warm-season hail outbreaks. Such a long-term trend in SHIP is driven both by thermodynamic and kinematic processes, represented by CAPE and $S_{06}$, respectively. From the perspective of SHIP, it appears that the interannual variability of hail outbreaks is dominated by kinematic processes, although thermodynamic processes still play a role. The interpretations of the long-term trends and interannual variability are reinforced when using seasonal accumulated SHIP.

WRs were used to investigate the link between large-scale circulations and hail outbreak activity. Relationships between WRs and hail outbreaks resemble those previously documented for tornado activity in Graber et al. (2025). WR-A is unfavorable for hail outbreaks, while WR-B and WR-D are favorable for hail outbreaks. Across all five WRs, positive and negative SHIP anomalies correspond well with positive and negative hail outbreak anomalies, respectively. Additionally, positive and negative SHIP anomalies correspond well with positive and negative anomalies of CAPE and $S_{06}$. This spatial alignment indicates that WRs modulate hail outbreak frequency through their influence on CAPE and $S_{06}$. While substantial work has linked WRs to tornado activity (Miller et al. 2020; Graber et al. 2025, 2026; Tippett et al. 2024), this appears to be the first study to link WRs to hail activity.

An empirical model based on WR frequency explains a meaningful portion of the interannual variability in hail outbreaks (cc=0.38, p=0.03). More specifically, it captures years with frequent hail outbreaks that are missed by SHIP alone, but it does not capture the positive long-term trend of hail outbreaks. This suggests that dynamical changes represented by WRs help explain the interannual variability of hail outbreaks. WRs influence the interannual variability of hail via the variability of CAPE and $S_{06}$ over the CONUS. In addition, modes of low-frequency variability, such as the El Niño Southern Oscillation, have been shown to modulate both WR frequency (Lee et al. 2023; Graber et al. 2026) and hail activity (Allen et al. 2015; Jeong et al. 2021). Future work should further investigate thermodynamic variables and the sources of predictability for hail outbreaks, including internal climate variability and external forcings.

*Acknowledgements*

Wang is supported by the NSF Grant 2518299. We acknowledge the NSF National Center for Atmospheric Research (NCAR) Computation and Information Systems Laboratory (CISL), sponsored by the National Science Foundation, for providing computing resources through Derecho (*doi:10.5065/qx9a-pg09*) and Casper (*https://ncar.pub/casper*). We acknowledge the Keeling computing cluster of the School of Earth, Society, and Environment (SESE) at the University of Illinois Urbana-Champaign. All ERA5 data used in this study are available at the research data archive and via the Copernicus Climate Change Service. Hail report data are available through the NOAA Storm Prediction Center severe weather database.

*Code Availability*

Code to create figures 1-3 is available at https://github.com/Matt0604/Hail_Trends_Graber. This link also contains code for the empirical model in figure 6. Relevant data files used to make the figures are also available.

*Data Availability Statement*

The ERA5 data are available at the NCAR research data archive (RDA) (d633000) and Copernicus Climate Data Store (CDS): https://doi.org/10.24381/cds.bd0915c6 (Hersbach et al. 2023a) and https://doi.org/10.24381/cds.adbb2d47 (Hersbach et al. 2023b). The hail report data are available through the NOAA Storm Prediction Center (SPC) Severe Weather Database:

https://www.spc.noaa.gov/wcm/#data. NCEI property and crop loss data is available at the Storm Events Database: https://www.ncei.noaa.gov/stormevents/details.jsp.

*Conflicts of Interest*

The contact author declares that none of the authors have any conflicts of interest.

## Supplementary Figures for

Seasonality of Hail Outbreaks in the United States and its link to weather regimes

Matthew Graber, Robert J. Trapp, & Zhuo Wang

Corresponding Author: Matthew Graber, mgraber2@illinois.edu

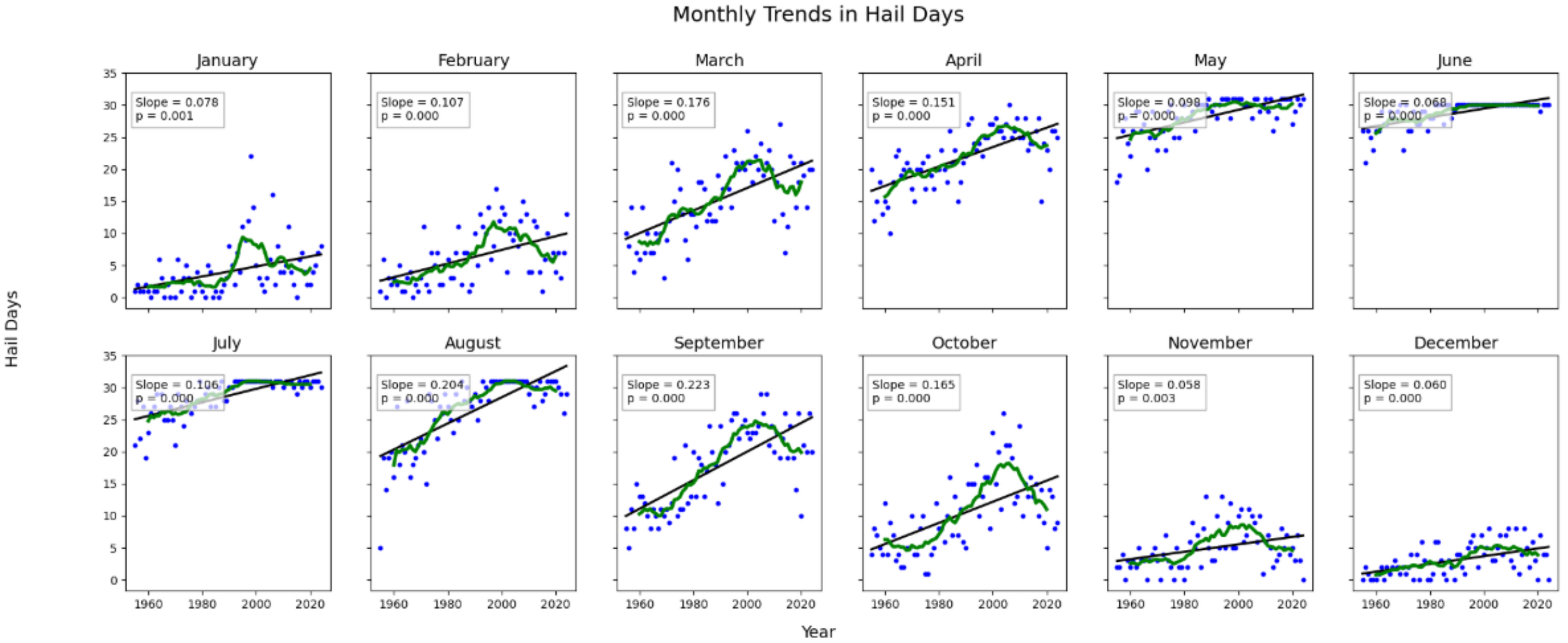

**Figure S1:** Severe hail days by month during 1960-2024 in CONUS (blue dots). Five-year running mean is indicated in green, and its linear trend is indicated by a dashed black line and summarized in each panel.

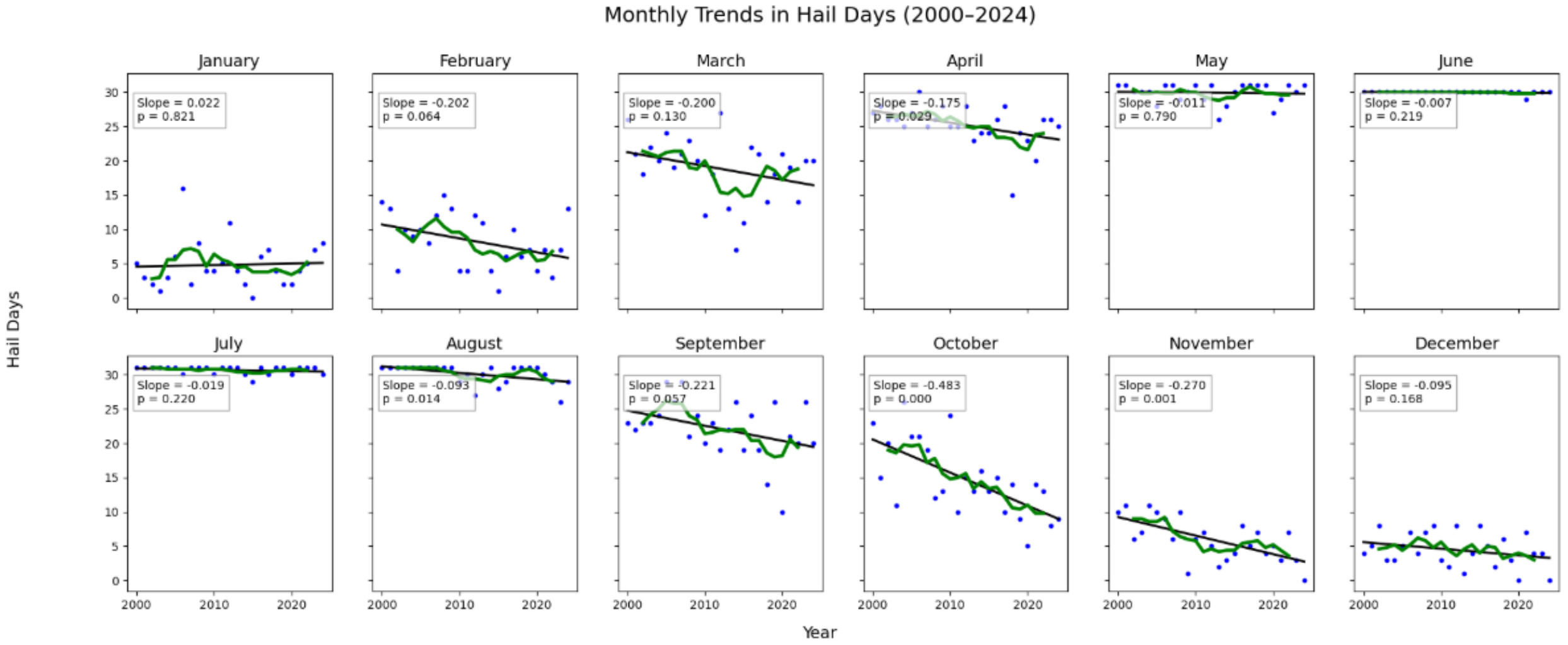

**Figure S2:** Severe hail days by month during 2000-2024 in CONUS (blue dots). Five-year running mean is indicated in green, and its linear trend is indicated by a dashed black line and summarized in each panel.

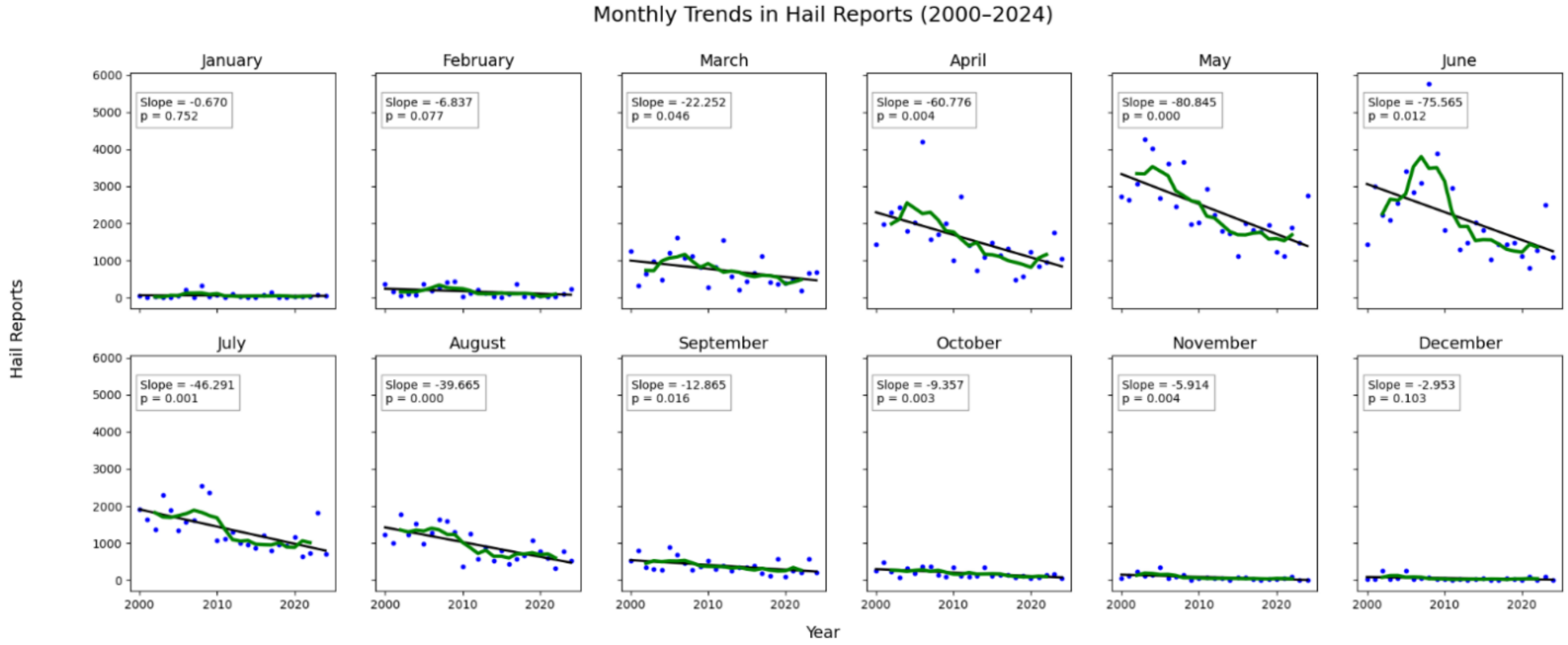

**Figure S3:** Severe hail reports by month during 2000-2024 in CONUS (blue dots). Five-year running mean is indicated in green, and its linear trend is indicated by a dashed black line and summarized in each panel.

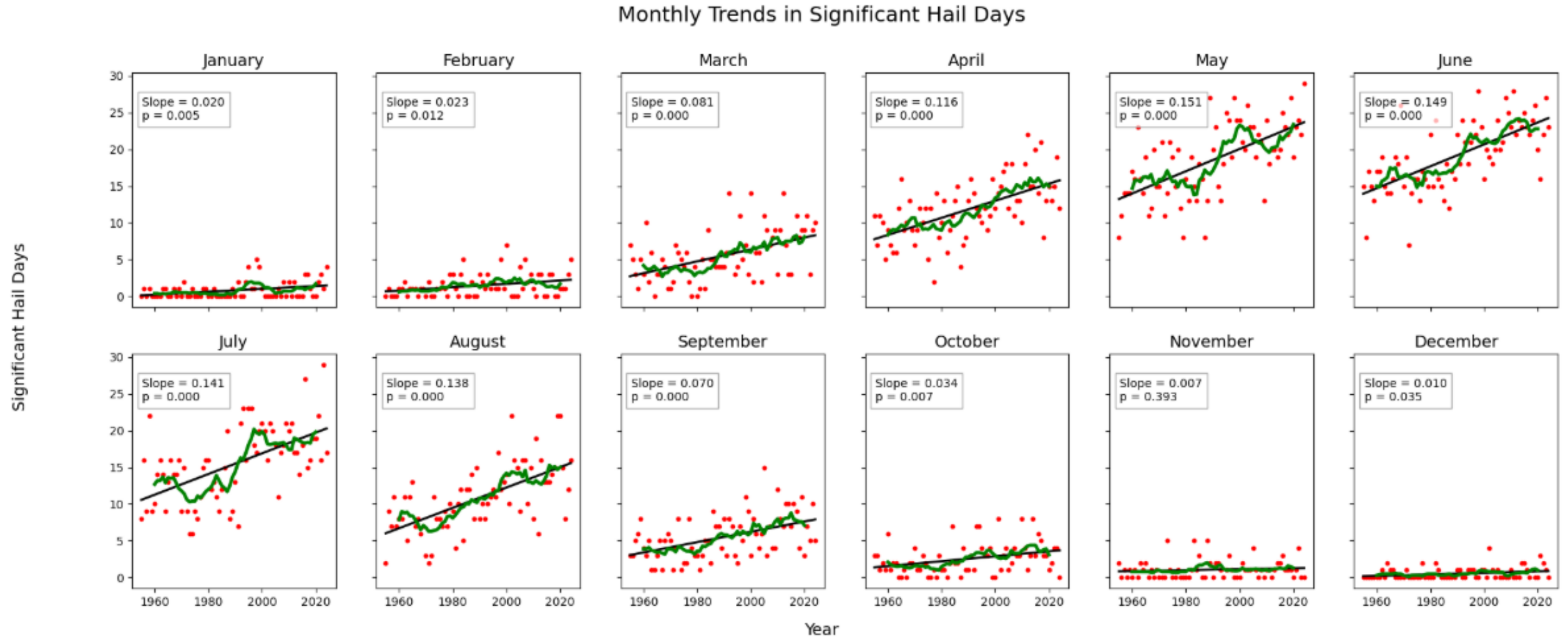

**Figure S4:** Significant hail days by month during 1960-2024 in CONUS (red dots). Five-year running mean is indicated in green, and its linear trend is indicated by a dashed black line and summarized in each panel.

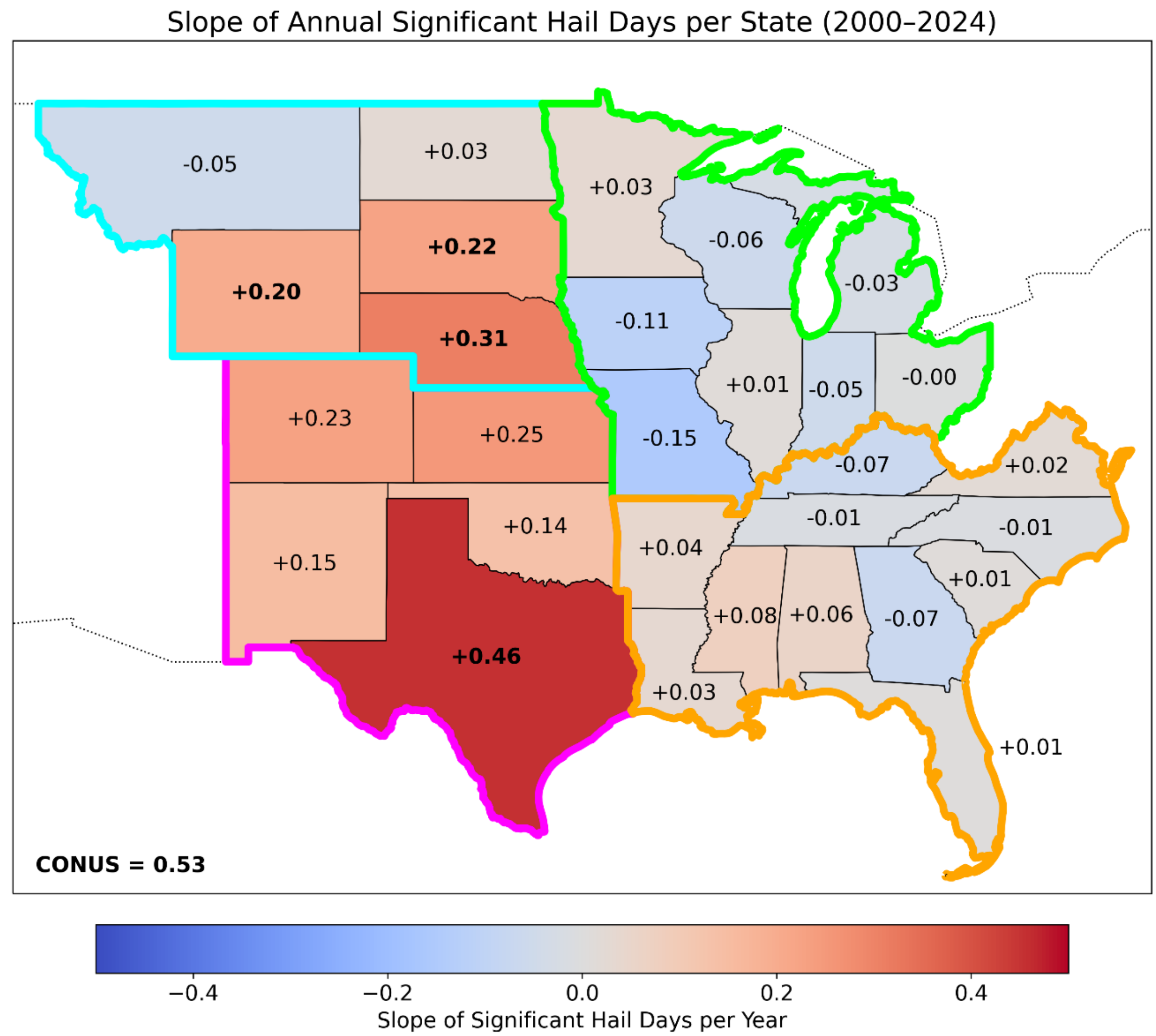

**Figure S5:** Slope of significant hail days per year by state for 2000-2024. Significant slopes are labeled in bold with the CONUS value in the bottom left. Regions are outlined in magenta (Southern Great Plains), cyan (Northern Great Plains), green (Midwest), and orange (Southeast).

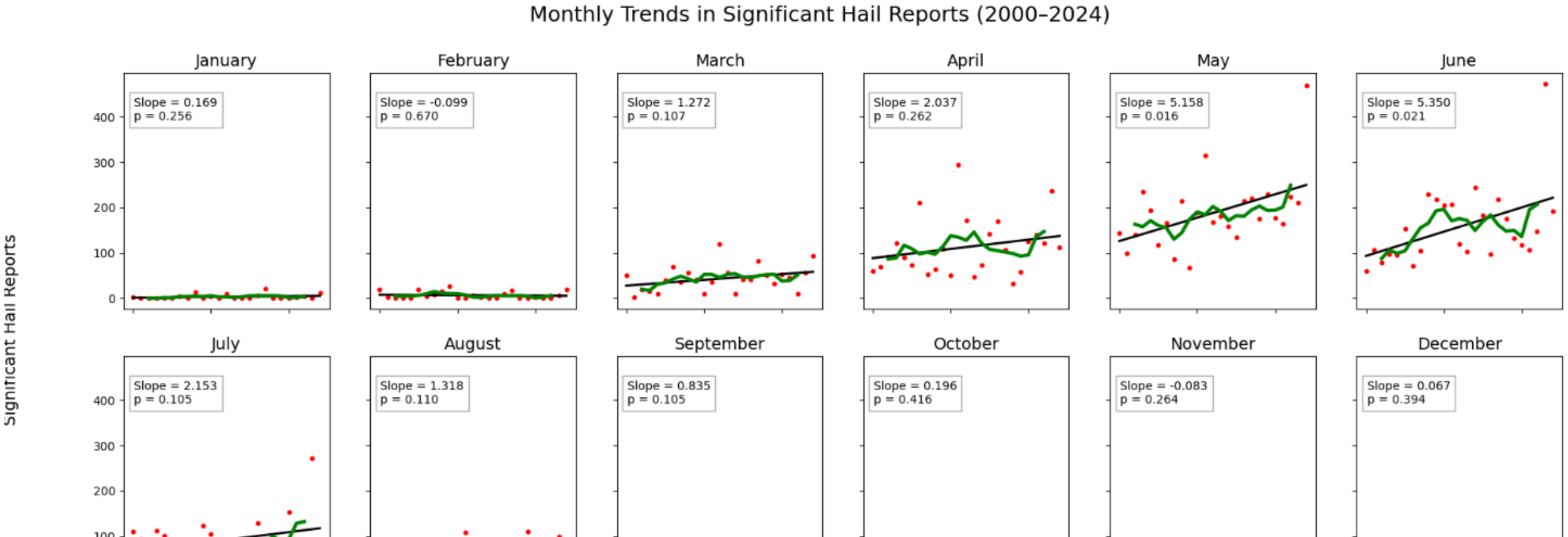

**Figure S6:** Significant hail reports by month during 2000-2024 in CONUS (red dots). Five-year running mean is indicated in green, and its linear trend is indicated by a dashed black line and summarized in each panel.

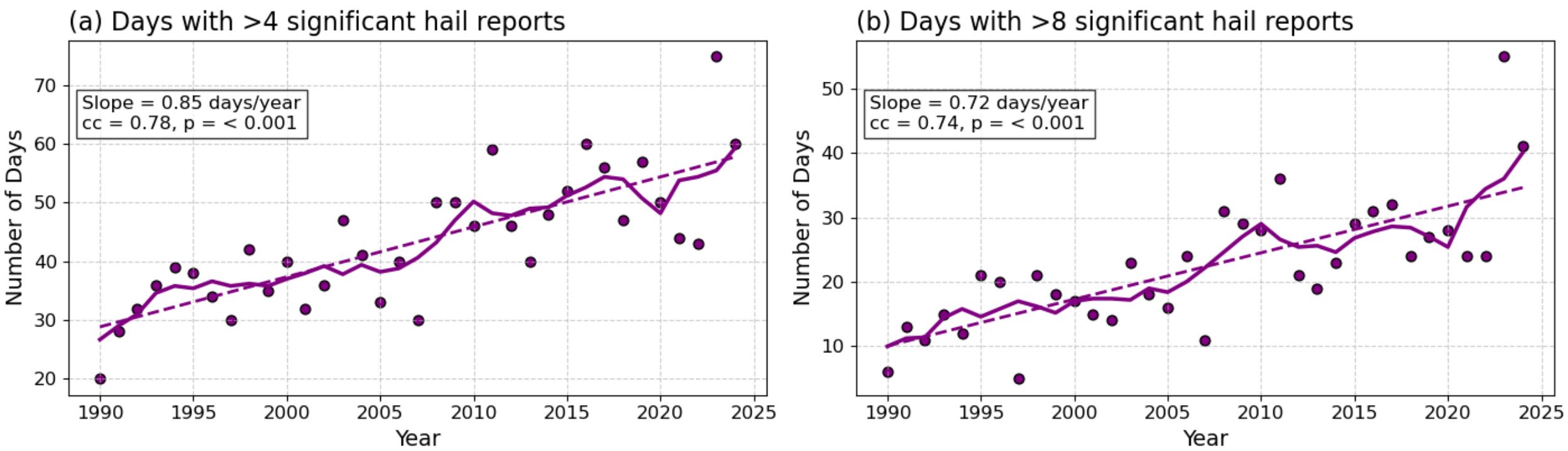

**Figure S7:** Days with (a) >4 and (b) >8 significant hail reports, with 5-year running means and linear regression statistics. The respective thresholds of >4 and >8 represent the 60th and 80th percentile values based on the number of daily significant hail reports.

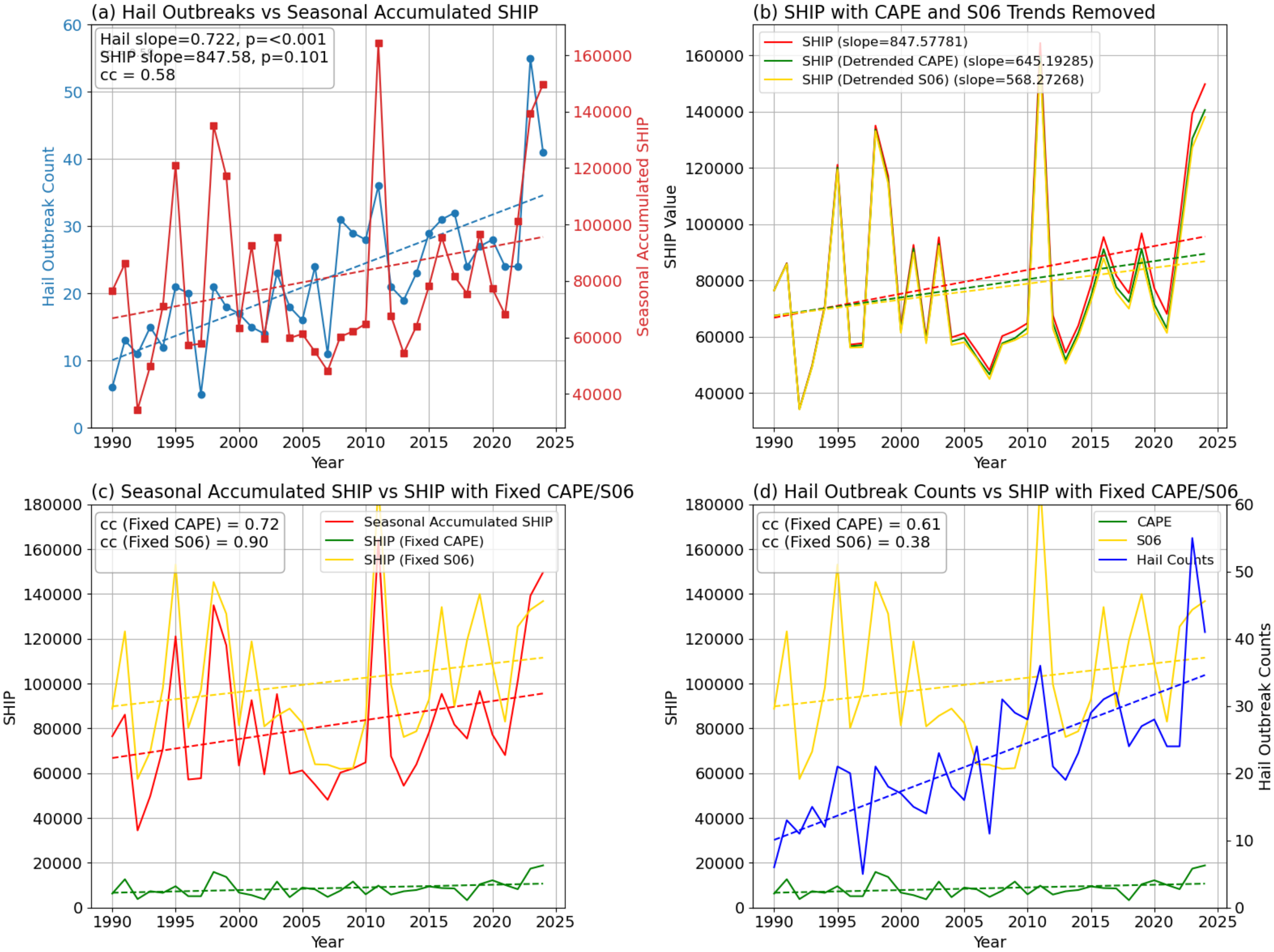

**Figure S8: (a)** Warm Season hail outbreak days (red) and warm season accumulated SHIP (where SHIP ≥ 1.0; blue/land grid points only) from 1990-2024 with linear regression statistics for each time series and the correlation coefficient between both time series. **(b)** Time series of SHIP (red), SHIP with CAPE long-term trend removed, and SHIP with $S_{06}$ long-term trend removed. **(c)** Warm-season accumulated observed SHIP (blue), and SHIP with fixed CAPE (green) and $S_{06}$ (yellow). Correlation coefficients are between the observed SHIP time series and the CAPE and $S_{06}$ time series **(d)** Warm-season hail outbreaks (red) vs SHIP with fixed CAPE (green) and $S_{06}$ (yellow) and corresponding correlation coefficients.

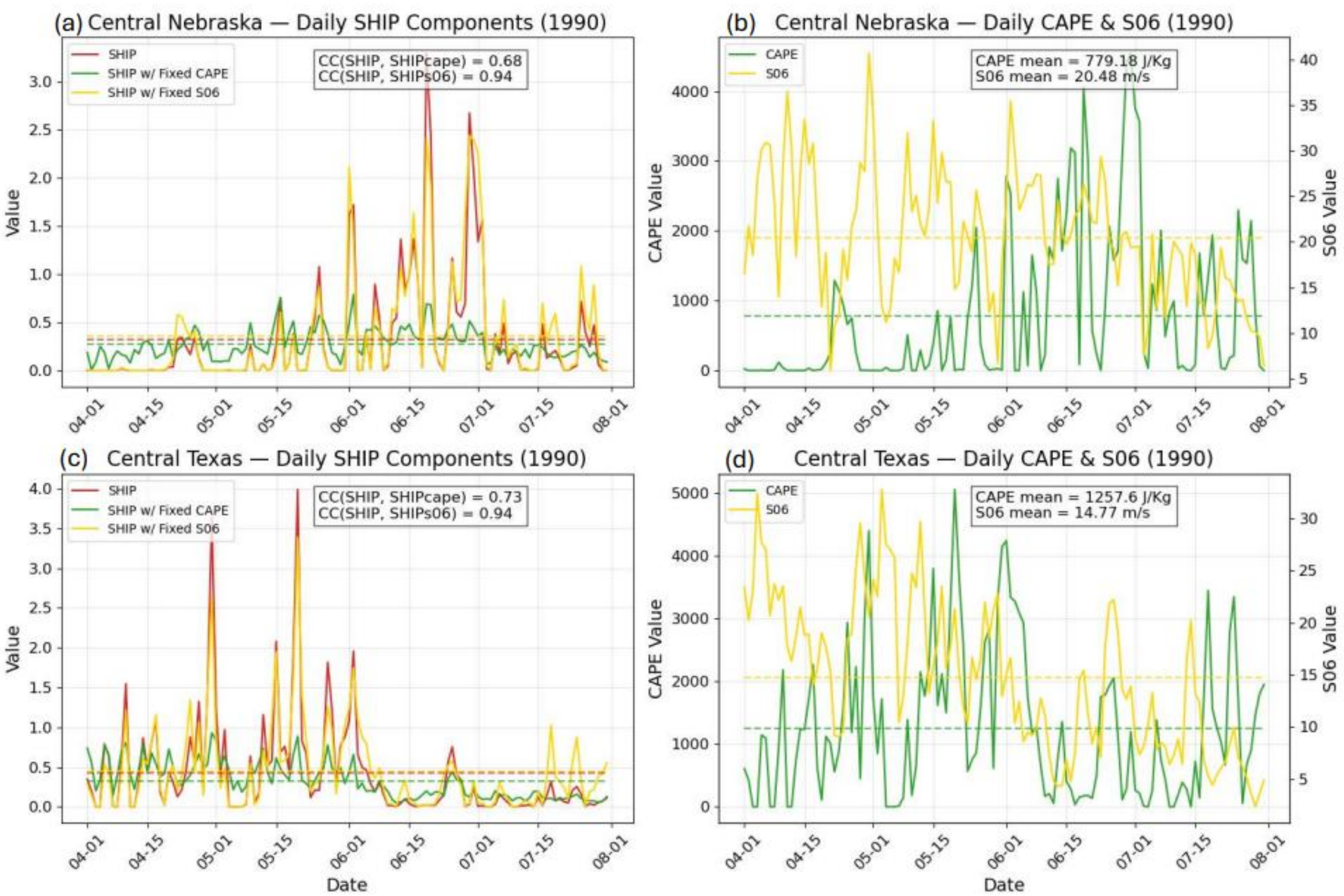

**Figure S9:** (a, c) 1990 observed daily mean SHIP (red) in central Nebraska (a) and central Texas (c) with corresponding daily mean SHIP time series with fixed CAPE (green) and fixed $S_{06}$ (yellow). (b, d) 1990 observed time series of daily max CAPE (green) and $S_{06}$ (yellow) in central Nebraska (b) and central Texas (d). Dashed, horizonal lines correspond to the seasonal mean.